\documentclass[aps,prd,preprintnumbers,nofootinbib]{revtex4}
\usepackage{amssymb}
\usepackage{bm}
\usepackage{enumerate}
\usepackage[font=normalsize]{subfig}
\usepackage{hyperref}

\usepackage{amsmath}
\usepackage{feynmf}
\usepackage{slashed}
\usepackage{graphicx}
\usepackage{mathtools}
\usepackage{color}
\newcommand\be{\begin{equation}}
\newcommand\ee{\end{equation}}

\newcommand\bea{\begin{eqnarray}}
\newcommand\eea{\end{eqnarray}}
\newcommand\cns{CE$\nu$NS }

\begin{document}
\begin{center}
\hfill MI-TH-1878
\end{center}
\title{Coherent Elastic Neutrino Nucleus Scattering (CE$\nu$NS) as a probe of $Z'$ through kinetic and mass mixing effects}
\author{Mohammad Abdullah$^{\bf a}$}
\author{James~B.~Dent$^{\bf b}$}
\author{Bhaskar~Dutta$^{\bf a}$}
\author{Gordon L. Kane$^{\bf c}$}
\author{Shu Liao$^{\bf a}$}
\author{Louis~E.~Strigari$^{\bf a}$}

\affiliation{$^{\bf a}$ Mitchell Institute for Fundamental Physics and Astronomy,
   Department of Physics and Astronomy, Texas A\&M University, College Station, TX 77845, USA}
\affiliation{$^{\bf b}$ Department of Physics, Sam Houston State University, Huntsville, TX 77341, USA}   
\affiliation{$^{\bf c}$ Leinweber Center  for Theoretical Physics, University of Michigan, Ann Arbor, MI 48109}

\begin{abstract}
We examine the current constraints and future sensitivity of Coherent Elastic Neutrino-Nucleus Scattering (CE$\nu$NS) experiments to mixing scenarios involving a $Z^\prime$ which interacts via portals with the Standard Model. We contrast the results against those from fixed target, atomic parity violation, and solar neutrino experiments. We demonstrate a significant dependence of the experimental reach on the $Z'$ coupling non-universality and the complementarity of \cns to existing searches.
\end{abstract}

\maketitle

\section{Introduction}

A straightforward approach to extending the Standard Model (SM) involves adding an extra $U(1)$ gauge group. Such extensions arise, for instance, in the context of Grand Unified Theories~\cite{Langacker:1980js} and string theory~\cite{Faraggi:1990ita} as well as in various phenomenological studies of Beyond Standard Model (BSM) physics such as hidden sectors \cite{Mizukoshi:2010ky} and solutions to B anomalies \cite{Gauld:2013qja}. Given such a gauge group with gauge field $Z'$ and field strength $F'_{\mu\nu}$, matter content that is charged under both this and a SM gauge group will generically lead to kinetic mixing of the form $\epsilon F'_{\mu\nu}F^{\mu\nu}$ where $F_{\mu\nu}$ denotes a SM field strength and $\epsilon$ parametrizes the strength of the mixing. This, in turn, can lead to $Z'$ interactions with SM fermions whose nature depends on the details of the model and can be explored by a variety of experiments. A similar result is obtained if a mass mixing with the SM $Z$ is instead generated. 
While both $\epsilon$ and the $Z'$ mass, $m_{Z'}$, vary depending on the model, much of the recent work has been focused on a range of $10^{-6}\lesssim \epsilon \lesssim 10^{-2}$ with $m_{Z'} \lesssim 10$ GeV.  

Studies of kinetic mixing have existed for decades \cite{Holdom:1985ag,delAguila:1988jz,Babu:1997st} (for a review of $Z'$ models in general see \cite{Langacker:2008yv}), with a tremendous number of experimental searches placing bounds on both the kinetic mixing parameter $\epsilon$, and the mass, $m_{Z'}$, of the new gauge field. A relatively new probe is neutral current neutrino scattering processes, in particular coherent elastic neutrino-nucleus scattering (CE$\nu$NS). The CE$\nu$NS process was predicted over forty years ago \cite{Freedman:1973yd}, and has long been discussed as a probe of the electroweak coupling structure~\cite{Barranco:2005yy,Barranco:2007tz}. It is very sensitive to alterations of the weak mixing angle~\cite{Akimov:2015nza,Cerdeno:2016sfi} such as those that arise with the introduction of hidden sector $Z'$ gauge bosons \cite{Bouchiat:2004sp,Dror:2017nsg}. 

\par The \cns process has only recently been experimentally detected by the COHERENT collaboration \cite{Akimov:2017ade}. COHERENT utilizes a proton beam on a fixed mercury target which produces charged pions that decay at rest to neutrinos, which can then scatter coherently in a CsI target. Recently, it has been shown that this measurement of CE$\nu$NS by the COHERENT collaboration can probe kinetic mixing scenarios where a $Z'$ that is produced via \emph{neutral} pion decay itself decays to dark matter particles that can then scatter coherently in the CsI target \cite{deNiverville:2015mwa,Ge:2017mcq} (the neutral pions are produced along with charged pions in the proton collisions with the fixed target).

One popular approach to constrain new physics via \cns involves placing bounds on Non-Standard Neutrino Interactions (NSI) in which otherwise unspecified high energy physics has been integrated out of the theory \cite{Dutta:2015vwa}\cite{Lindner:2016wff}. Such limits only apply to mediators with masses sufficiently above the momentum transfer scale of the  interaction. Some studies have placed bounds on simplified models including a light $Z'$ \cite{Liao:2017uzy,Dent:2016wcr,Kosmas:2017tsq,Barranco:2007tz,Lindner:2018kjo}, but they all assume a direct coupling to both neutrinos and first generation quarks rather than indirect couplings induced from mixing.

In this paper we examine the sensitivity of present and future measurements of \cns to two scenarios where the mixing is induced at high energy resulting in a $Z'$ that couples similarly to the hypercharge gauge boson (dark hypercharge boson) or to the $Z$ boson (dark $Z$ boson), as well as a scenario where the mixing is generated at low energy by couplings to second and third generation leptons. Such scenarios have the feature that the underlying physics dictates specific patterns of $Z'$ couplings to the SM fermions which influences the experimental reach of different experiments. We compare the \cns sensitivity to the sensitivity obtained by fixed target, atomic parity violation and solar neutrino experiments, and find the \cns limits to be competitive and in some cases stronger.

In addition to further measurements by the COHERENT collaboration, there exists a suite of near-term experimental programs designed to measure CE$\nu$NS with high statistics at nuclear reactor facilities \cite{Agnolet:2016zir,Billard:2016giu,Aguilar-Arevalo:2016khx,Strauss:2017cuu} with ultra-low threshold detectors. We find the projected reach of cryogenic germanium and silicon detectors at such reactor facilities to be stronger than future COHERENT measurements for $Z'$ masses in the sub-10 MeV range. 

The outline of this paper is as follows: In Sec.~\ref{sec:cenns} we review the CE$\nu$NS process, in Sec.~\ref{sec:mixing} we present the details of kinetic and mass mixing, in Sec.~\ref{sec:bounds} we briefly outline the existing bounds, in Sec.~\ref{sec:setup} we explain our numerical setup, in Sec.\ref{sec:results} we present the results from current and projected measurements, and we summarize our results in Sec.~\ref{sec:discussion}.

\section{Coherent Elastic Neutrino-Nucleus Scattering}
\label{sec:cenns}

\cns occurs whenever the momentum transfer between the neutrino and the nucleus is smaller than or comparable to the inverse size of the nucleus, which takes place for incident neutrino energies of $E_{\nu} \lesssim\mathcal{O}(50)$ MeV. For such low energies, the neutrino effectively ``sees" the entire nucleus rather than the individual components, leading to an enhancement in the scattering cross section that scales approximately as the square of the number of neutrons. The neutron number is dominant because the proton coupling that contributes to the \cns process is more than an order of magnitude smaller than the neutron coupling. 

In the SM, the differential cross-section for a neutrino scattering off of
a target electron or quark of mass $m$ through a $Z$ exchange is
\be
\frac{d\sigma}{dE_R} = \frac{G_F^2 m}{2\pi}\left((g_v+g_a)^2 +
(g_v-g_a)^2\left(1-\frac{E_R}{E_{\nu}}\right)^2 +
(g_a^2-g_v^2)\frac{mE_R}{E_\nu^2}\right)
\,,
\label{eq:SM}
\ee
where $G_F$ is the Fermi constant, $E_R$ is the recoil energy, $E_\nu$ is the incident neutrino energy,  $(g_v,g_a)= (T_3-2Q_{\rm em}{\rm{sin}}^2\theta_W,T_3)$ are the vector and axial-vector couplings to the $Z$-boson, $T_3$ is the third component of the weak isospin,
$Q_{\rm em}$ is the electromagnetic charge, and $\theta_W$ is the weak mixing angle ($T_{3e} = -1/2$ in our convention).
 There is also a contribution from neutrinos scattering off electrons
due to the  the charged-current $t$-channel exchange of a $W$-boson.  In order to account for the full momentum dependence of the nuclear scattering interaction, the differential cross-section must be multiplied by a form factor. In the present work we use the standard Helm form factor \cite{Helm:1956zz}.

The SM cross-section above can be modified if there exists a new
mediating particle which couples to neutrinos and either electrons or quarks.  Let us consider a new vector particle $Z_{\mu}'$ with the following interaction terms in the Lagrangian:
\be
\mathcal{L} \supset Z_{\mu}'(g_{\nu}'\bar{\nu}_L\gamma^{\mu}\nu_L + g_{f,v}'\bar{f}\gamma^{\mu}f + g_{f,a}'\bar{f}\gamma^{\mu}\gamma^5 f),
\label{eq:Zprime}
\ee
where $g_{\nu}'$, $g_{f,v}'$, and $g_{f,a}'$ are constants associated with new physics. The effects of this new field can be accommodated by a redefinition of the couplings in Eq.~(\ref{eq:SM}):
\be
(g_v,g_a) \,\Rightarrow\, ( g_v, g_a )
+ \frac{g_{\nu}'\, (g_{f,v}',\pm g_{f,a}')}{\sqrt{2}G_F\, (q^2 + M_{Z'}^2)}),
\label{eq:SMmod}
\ee
where $q^2$ is the momentum trasnfer and the $(-)$ sign applies for the case of antineutrino scattering. Although only new vector mediators are considered here, one can introduce other types of mediators and associated couplings, e.g., the mediator could be of scalar and/or  pseudoscalar type~\cite{Cerdeno:2016sfi,Lindner:2016wff,Dent:2016wcr}.

To-date the \cns process has only been measured by the COHERENT collaboration. The dominant portion of the neutrino source is a prompt flux of $\nu_{\mu}$ of energy 30 MeV from $\pi^+ \rightarrow \mu^+ + \nu_{\mu}$ decays, with sub-dominant components from delayed $\nu_e$ and $\bar{\nu}_{\mu}$ fluxes originating from the $\mu^+$ decay (the form of the energy distributions for these neutrino beams are given, for example, in \cite{Coloma:2017egw}). The first detection of the \cns process was obtained with a 14.6 kg array of CsI scintillators with a 4.25 keV detection threshold. The COHERENT collaboration plans to implement a ton-scale liquid argon detector and a ton-scale array of NaI scintillators \cite{Akimov:2017ade} which will be used for the projections in the current work.

Nuclear reactor facilities aim to detect the \cns process with a $\bar{\nu}_e$ flux with energies $\sim 1$ MeV. These neutrinos will produce sub-keV nuclear recoils, necessitating the use of low threshold detector technology such as that developed for dark matter direct detection experiments. The \cns scattering rate per target mass at the proposed reactor experiments is projected to be a few orders of magnitude greater than that measured by the COHERENT experiment. This is due to reactor neutrino fluxes which are roughly 5-6 orders of magnitude greater than that of COHERENT. However this flux advantage is mitigated because the scattering rate is proportional to the square of the incident neutrino energy. With reactors producing a neutrino source of roughly 20 times smaller energy than at COHERENT, this will relatively reduce the scattering rate by a factor of about 400. The projected ton-scale targets at COHERENT will help to partially offset this comparative rate deficiency, as reactor experiments expect to deploy detectors of total mass of $\mathcal{O}(10)$ kg. For the projections in the present work we will utilize cryogenic Ge and Si with 100 eV nuclear recoil thresholds, and a total exposure of 100 kg$\cdot$yr.

\section{Kinetic and Mass Mixing}
\label{sec:mixing}

We will now examine how such couplings could arise in mixing scenarios first examined in~\cite{Holdom:1985ag}. Let us consider the SM hypercharge gauge group $U(1)_Y$ with gauge field $B$ and a dark abelian gauge group $U(1)_X$ with gauge field $X$ and suppose we have the following gauge kinetic terms
\be L_{gauge}=-{1\over{4}}F^{\mu\nu}_aF_{a\mu\nu}-{1\over{4}}F^{\mu\nu}_bF_{b\mu\nu}-
{\epsilon\over{2}}F^{\mu\nu}_aF_{b\mu\nu}\label{lagrangian}\ee
where $\epsilon$ parameterizes the mixing of the two $U(1)$s.  The mixing can be generated by a loop of heavy fields charged under both groups. For fields with masses $m$ and $m'$ and gauge couplings $g_Y$ and $g_X$ we get $\epsilon \sim {{g_Y g_X}\over{12 \pi^2}}\log{m'^2\over {m^2}}$. For $\mathcal{O}(1)$ couplings and a mass difference of a factor of 10 or less, this restricts us in the range $\epsilon\,\lesssim\,10^{-2}$. One can remove the mixing term by a field redefinition: $B_\mu\rightarrow B_\mu+\epsilon X_\mu$ which induces new couplings of $X$ to the SM fermions. Note that this leads to non-diagonal terms in the $Z-X$ mass matrix. Such mass-mixing can be controlled by introducing additional mass mixing from an extended Higgs sector of whose details we remain oblivious ~\cite{Gopalakrishna:2008dv,Davoudiasl:2012ag}. 

Accounting for the diagonalization of both the kinetic and mass terms, as well as the SM electroweak rotation, the mass eigenstates $A$, $Z$ and $Z^\prime$ can be  expressed in terms of $B_\mu$, $W^3_\mu$ and $X_\mu$ by the following transformation matrix:
 \be
 \left(\begin{array}{c}   
    B_\mu\\
       W^3_\mu \\
       X_\mu
         \end{array}\right)= \left(\begin{array}{ccc}   
  \cos{\theta_w} & -\epsilon\, \sin\alpha-\sin{\theta_w}\cos\alpha&\sin{\theta_w}\sin\alpha-\epsilon\, \cos\alpha\\
       \sin{\theta_w} & \cos{\theta_w} \cos\alpha & -\cos{\theta_w}\sin\alpha \\
       0& \sin\alpha & \cos\alpha
         \end{array}\right) \left(\begin{array}{c}   
   A_\mu\\
      Z_\mu \\
       Z'_\mu
         \end{array}\right),\ee
         
where the angle $\alpha$, which has implicit dependence on $\epsilon$, controls the $Z-Z'$ mass mixing.  $\alpha$   can arise from an extended Higgs sector  with a new symmetry breaking scale different from the EW scale. The extended Higgs sectors  can emerge from grand unified theory models~\cite{Langacker:1980js,Hewett:1988xc} and $\alpha$ can also arise via the Stueckelberg mechanism ~\cite{Feng:2014cla}. Note that the photon remains massless to all orders in the mixing parameters.

From this we can infer the $Z^\prime$-fermion-antifermion coupling to be:
\be
{-i g\over{\cos{\theta_w}}}\left[\cos\alpha(\tan\alpha-\epsilon\, \sin{\theta_w})\right]\left[T^3_L-{(\tan\alpha-\epsilon\,/\sin{\theta_w})\over{\tan\alpha-\epsilon\, \sin{\theta_w}}}\sin^2{\theta_w}Q\right],\ee

where $g$ is the SM $SU(2)$ gauge coupling. If we assume $\alpha$ to be small and define $\epsilon_B\equiv \cos\alpha(\tan\alpha-\epsilon\, \sin{\theta_w})$, the coupling becomes proportional to the hypercharge, and we refer to $Z^\prime$ as a dark hypercharge gauge boson. Explicitly, the coupling is given by: 
\be {i g\,{\tan{\theta_w}}}\,(Y_f/2)\,\epsilon_B,
\label{newcoupling}
\ee 
where $Y_f$ is the hypercharge of the SM fermion $f$. If, instead, we choose $\epsilon$ to be zero and define $\epsilon_z\equiv \sin\alpha$, then the $Z^\prime$ coupling reduces to 
\bea
\label{eq:darkz}
{-i g\over{\cos{\theta_w}}}\,\epsilon_z\,\left[T^3_L-\sin{\theta^2_w}Q\right],
\eea
and we call it a dark $Z$ boson. 

The case usually referred into in the literature as the dark photon corresponds to setting $\tan\alpha=\epsilon\, \sin{\theta_w}$ and will not be discussed in this work since it generates no couplings between the $Z'$ and neutrinos.

Another interesting possibility for probing new physics models is when the SM is extended with a non-universal $U(1)$ gauge symmetry
associated with $U(1)_{L_\mu-L_\tau}$. This symmetry has been discussed in various contexts including the flavor structures of neutrinos \cite{He:1991qd,He:1990pn}, lepton flavor violating Higgs decays \cite{Heeck:2014qea},  dark matter, and the recently reported flavor non-universality in B decays \cite{Altmannshofer:2016jzy}. This symmetry leads to interactions in the Lagrangian of the form:
$\mathcal{L}_{int} \supset g_{Z^\prime}Q_{\alpha\beta}(\bar{l}_\alpha\gamma^\mu l_\beta+\bar{\nu}_{L\alpha}\gamma^\mu \nu_{L\beta})Z^\prime_{\mu}$,
where, as before, $Z^\prime$ is the new gauge boson, $g_{Z^\prime}$ is the new gauge coupling, and $Q_{\alpha\beta}={\rm{diag}}(0,1,-1)$ gives the $U(1)_{L_\mu-L_\tau}$ charges. It is possible to extend this symmetry to the quark sector as well.  

At low energies, muon and tau loops generate kinetic mixing between the SM photon and $Z'$ of strength $\epsilon\propto (8e g_{Z^\prime})/(48 \pi^2)\text{log}(m_\tau/m_\mu)$ \cite{Kamada:2015era,Araki:2017wyg} (the $\mu$ and $\tau$ leptons can be replaced by second and third generation quarks if the symmetry is also extended to the second-third generation quark sector). Since this is generated at low energy, the diagonalization is done after electroweak symmetry breaking and results in a $Z'$ coupling to the first generation quarks equal to $\epsilon\,Q$. While this mixing is suppressed by a loop factor, this is compensated for by the direct coupling to neutrinos in $\nu_\mu$ scattering experiments.

A diagrammatic representation of the three scenarios is shown in Fig.(\ref{ScattDiag}). Fig. 1(a) is associated with the dark  $Z$ and dark hypercharge cases where the blobs contain the high energy physics responsible for the mixing $\epsilon$. Fig 1(b) corresponds to the $L_\mu-L_{\tau}$ case where the new gauge boson provides a direct coupling to a muon neutrino but communicates with first generations quarks through lepton loops. ($\nu_{\mu}$ or $\bar{\nu}_{\mu}$ are the relevant particles for the CE$\nu$NS process generated from the decay of charged pions, such as utilized by the COHERENT collaboration).

\section{Existing Bounds}
\label{sec:bounds}

\par Several experiments have placed bounds on the mixing parameters introduced above. In this section we review these bounds and their applicability to the mixing scenarios in consideration. 

In addition to the bounds below, our model can, in principle, be constrained by the meson decays $B\rightarrow K Z^{\prime}$ and $K\rightarrow \pi Z^{\prime}$. However, since  the amplitudes for these decay modes depend on the details of the Higgs sector, we are neglecting constraints from these decay modes in our analysis.

\subsection{Fixed target experiments}

One of the most stringent constraints on a light $Z'$ that couples to electrons is through fixed target experiments, also known as beam dump experiments. An electron beam is aimed at a high density target and the scattering products are observed. In such a process the scattered electron may emit a $Z'$ in a similar fashion as bremsstrahlung which can then decay to observable SM particles. This has been searched for by the SLAC E137 (20 GeV) \cite{Bjorken:1988as}, SLAC E141 (9 GeV) \cite{Riordan:1987aw}, and Fermilab E774 (275 GeV) \cite{Bross:1989mp} experiments and no candidate events have been observed. Another way of utilizing fixed target experiments that would constrain our models is by looking at pion decays ($\pi^0 \rightarrow \gamma Z' ,\; Z' \rightarrow e^+e^-$). This search has been performed by the NA48/2 Collaboration at CERN \cite{Batley:2015lha}\footnote{We thank the authors of \cite{Ilten:2018crw} for pointing this out and providing us with their data.}. We adopt the analysis of Refs. \cite{Ilten:2018crw,Bjorken:2009mm,Harnik:2012ni} to place limits on our models, and since they assume a coupling of the form $\epsilon e A^{\prime}_{\mu} J^{\mu}_{EM}$, the limits on the dark $Z$ and dark hypercharge require the following scalings respectively:

\begin{align}
\epsilon &\rightarrow \frac{e\cos\theta_w\epsilon}{g\sqrt{\left(-0.5+2\sin^{2}\theta_w\right)^2+\left(-0.5\right)^2}}\\
\epsilon &\rightarrow \frac{e\epsilon}{g_{Y}\sqrt{\left(-0.5\right)^{2}+\left(-1.5\right)^{2}}}
\end{align}

The advantage that these experiments have is that the limits are independent of the $Z'$ coupling to nucleons. On the flip side, the limits are very weak for the $L_\mu-L_\tau$ scenario since the couplings to both electrons and light quarks are suppressed by a loop factor. 

The $L_\mu-L_\tau$ scenario can be tested via the so-called neutrino trident production as first done in \cite{Altmannshofer:2014cfa}. In the SM, a neutrino beam may scatter inelastically off a fixed nuclear target producing a pair of muons at tree level which are then detected ($\nu N \rightarrow \nu N\mu^{+}\mu^{-}$). This process has been observed by both the CHARM-II collaboration \cite{Geiregat:1990gz} and the Columbia-Chicago-Fermilab-Rochester (CCFR) collaboration \cite{Mishra:1991bv} using muon-flavored neutrinos and found to be in agreement with the SM prediction. A $Z'$ which couples to muon flavor leptons should generate a similar signal and can therefore be constrained. For our analysis we reproduce the analytical approximations used in \cite{Altmannshofer:2014pba} to translate the bounds from the CCFR collaboration which finds:
\begin{equation}
\sigma_{CCFR}/\sigma_{SM} = 0.82 \pm 0.28
\end{equation}.


\subsection{Solar neutrinos} 

\par
\begin{figure}[h]
\centering
\subfloat[]{\includegraphics[height = 5cm]{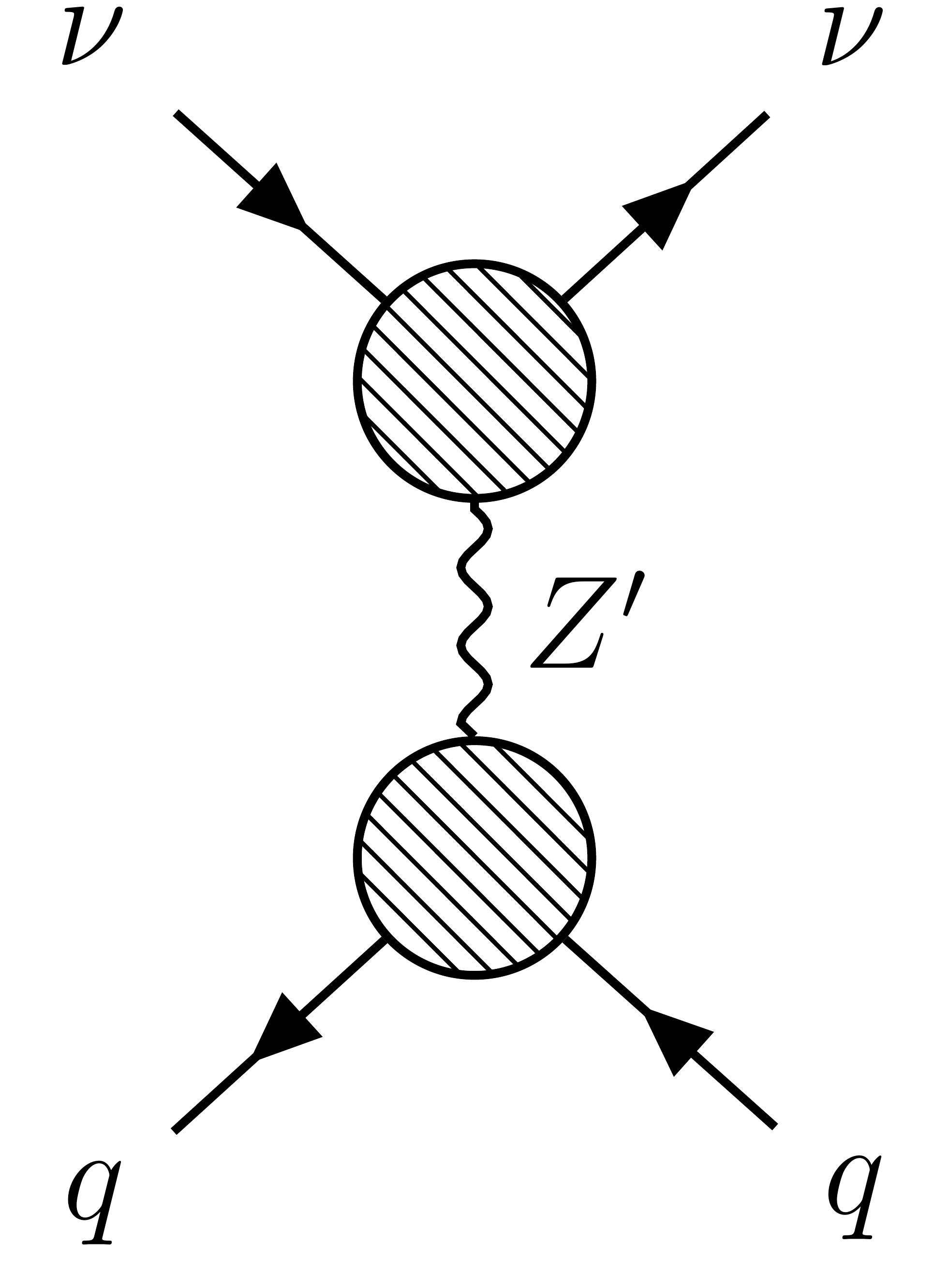}}
\hspace{0.5cm}
\subfloat[]{\includegraphics[height = 5cm]{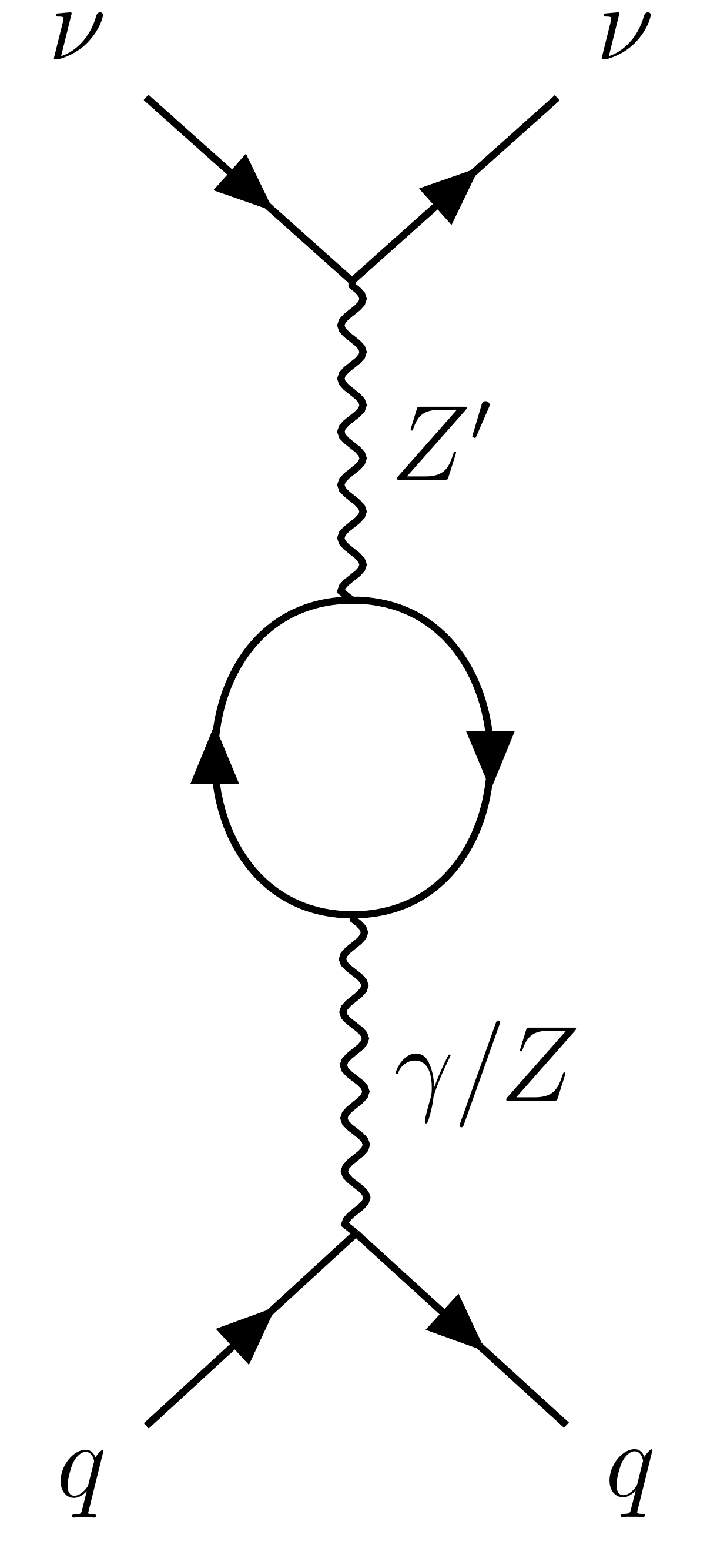}}
\caption{Neutrino nucleus scattering diagrams. Fig. 1(a) is for dark hypercharge and dark $Z$ bosons and Fig. 1(b) is for the $L_\mu-L_{\tau}$ model. }
\label{ScattDiag}
\end{figure}

 The elastic scattering of low-energy solar neutrinos provides a probe of $U(1)_{L_\mu-L_\tau}$ gauge bosons. In particular, the $U(1)_{L_\mu-L_\tau}$ gauge boson can contribute to neutrino-electron scattering from solar neutrinos through a loop induced coupling as shown in Fig.(\ref{ScattDiag}). The lowest energy elastic scattering data comes from the Borexino solar neutrino experiment. Borexino has now attained a $\sim 2.7\%$ uncertainty on the measurement of the low energy solar neutrino rate from the monotonic $^7$Be component of the solar flux, which are produced at 0.861 MeV~\cite{Agostini:2017ixy}. This improves previous measurements of this component by more than a factor of two. Further, Borexino also has strong constraints on other low energy components of the solar flux (pp, pep, and ${}^8$B). Using earlier Borexino $^7$Be data~ \cite{Bellini:2011rx}, Refs.~\cite{Kamada:2015era,Araki:2017wyg} calculate the contribution of a $U(1)_{L_\mu-L_\tau}$ gauge boson to this process, and generate parameter constraints by converting existing bounds \cite{Harnik:2012ni} on $U(1)_{B-L}$ models. 
In our analysis below, we have used the recently-released Borexino measurement of ${}^7$Be neutrinos \cite{Agostini:2017ixy} to update the bounds from solar neutrino scattering.

{{\subsection{$\Upsilon$ decays} 

Decays of $\Upsilon$ mesons to a photon plus a dark $Z$ or hypercharge boson, with the dark boson decaying further to $\mu^+\mu^-$, are constrained by BaBar results~\cite{Essig:2009nc}. This constraint becomes relevant for $M_{Z^\prime}\sim 10$ GeV. }}

\subsection{Atomic parity violation} 

Another powerful probe for our models comes from atomic parity violation. The $6S_{1/2}-7S_{1/2}$ nuclear transition in ${}^{133}$Cs is only allowed by the parity violation furnished by the electroweak force and has been measured by multiple collaborations to great precision \cite{Porsev:2009pr}\cite{Porsev:2010de}\cite{Wood:1997zq}\cite{Bennett:1999pd}. The result can be expressed as a value of the nuclear weak charge defined as $Q_w = -N+Z(1-4 \sin^2\theta_w)$ and compared to the SM prediction \cite{Marciano:1982mm}\cite{Marciano:1983ss}\cite{Marciano:1990dp}:
\begin{align}
Q^{SM}_w({}^{133}_{55}Cs) &= -73.16(5)\\
Q^{exp}_w({}^{133}_{55} Cs) &= -73.16(35)
\end{align}.

Both the dark $Z$ and dark hypercharge scenarios lead to parity violating couplings to the SM fermions which would modify the values of the electroweak parameters and hence $Q_w$. The agreement with the SM then leads to strong limits on $\epsilon$. We utilize the results presented in \cite{Davoudiasl:2012ag} using 30 MeV as the energy scale of the measurement \cite{Porsev:2010de}. Note that, once again, these limits are highly suppressed in the $L_\mu - L_\tau$ scenario due to the loop factor.


 \section{Numerical Setup}
 \label{sec:setup}
 
 To evaluate the current and projected sensitivity of future \cns experiments, we use a $\chi^2$ method to calculate the bounds on $\epsilon$ at the $2\sigma$ confidence interval.
 Following \cite{Liao:2017uzy} we define:
 \begin{equation}
\chi^{2}=\sum_{\mathrm{bins,detectors}}\frac{\left(N_{exp}-\left(1+\beta\right)N_{pred}\right)^{2}}{N_{bg}+N_{exp}}+\left(\frac{\beta}{\sigma_{\beta}}\right)^{2}
\label{chi}
\end{equation}
where $N_{exp}$ is the expected number of events in the SM (or the observed number of events in the case of current COHERENT limits), $N_{pred}$ is the number of predicted events in our model, $N_{bg}$ is the number of background events, $\sigma_\beta$ is the fractional systematic uncertainty, and $\beta$ is the corresponding nuisance parameter. 
In our calculation $\sigma_{\beta} = 0.1$ \cite{Akimov:2017ade}.
We scan over the range 1 MeV $\leq \,M_{Z'}\,\leq$ 10 GeV and set limits at the $2\sigma$ level, which means that any point above our exclusion curves can be interpreted to be within discovery sensitivity with at least $95\%$ probability.

The data used in analysis includes both current data and future reactor and accelerator projections. 
The current data set consists of the observed number of events per bin from COHERENT \cite{Akimov:2017ade}. 
Future projections are set by the Asimov dataset for a NaI/Ar detector for accelerator experiments and Ge/Si for reactor experiments. 
The Asimov dataset is the set of simulated data which will cause the likelihood to be maximized at the expected points of all the parameters \cite{Cowan:2010js}.
For future reactor experiments, we assume an exposure of $100\, \mathrm{kg \times year}$, while for future accelerator experiments we assume $1 \, \mathrm{ton \times year}$ and $10\, \mathrm{ton \times year}$ of exposure.
The background is taken to be $1~\mathrm{dru}$ (1 dru = 1 event/kg/day/keV) for Ge and Si detectors (reactor experiments) and $5\times10^{-3}~\mathrm{dru}$ for NaI and Ar detectors (COHERENT).
We choose 20 energy bins in our analysis to account for the shape information in the energy spectrum.
The detection threshold used in our calculation is $100\,\mathrm{eV}$ for Ge and Si detectors and $2\, \mathrm{keV}$ for NaI and Ar detectors.
The reactor neutrino flux is assumed to be $1.5\times10^{12}/\left(\mathrm{cm^2\cdot s}\right)$, typical of a $1\,\mathrm{MW}$ reactor at $1\,\mathrm{m}$ from the detector or, equivalently, a $1\,\mathrm{GW}$ reactor $30\,\mathrm{m}$ away from the detector. The energy distribution is taken from \cite{Vogel:1989iv}. For the accelerator experiment we use a flux of $4.29\times10^9/\left(\mathrm{cm^2\cdot s}\right)$ \cite{Akimov:2017ade} with the energy distribution given in \cite{Coloma:2017egw}.

\section{Results}
\label{sec:results}

\begin{figure}[h]
\includegraphics[scale = .7]{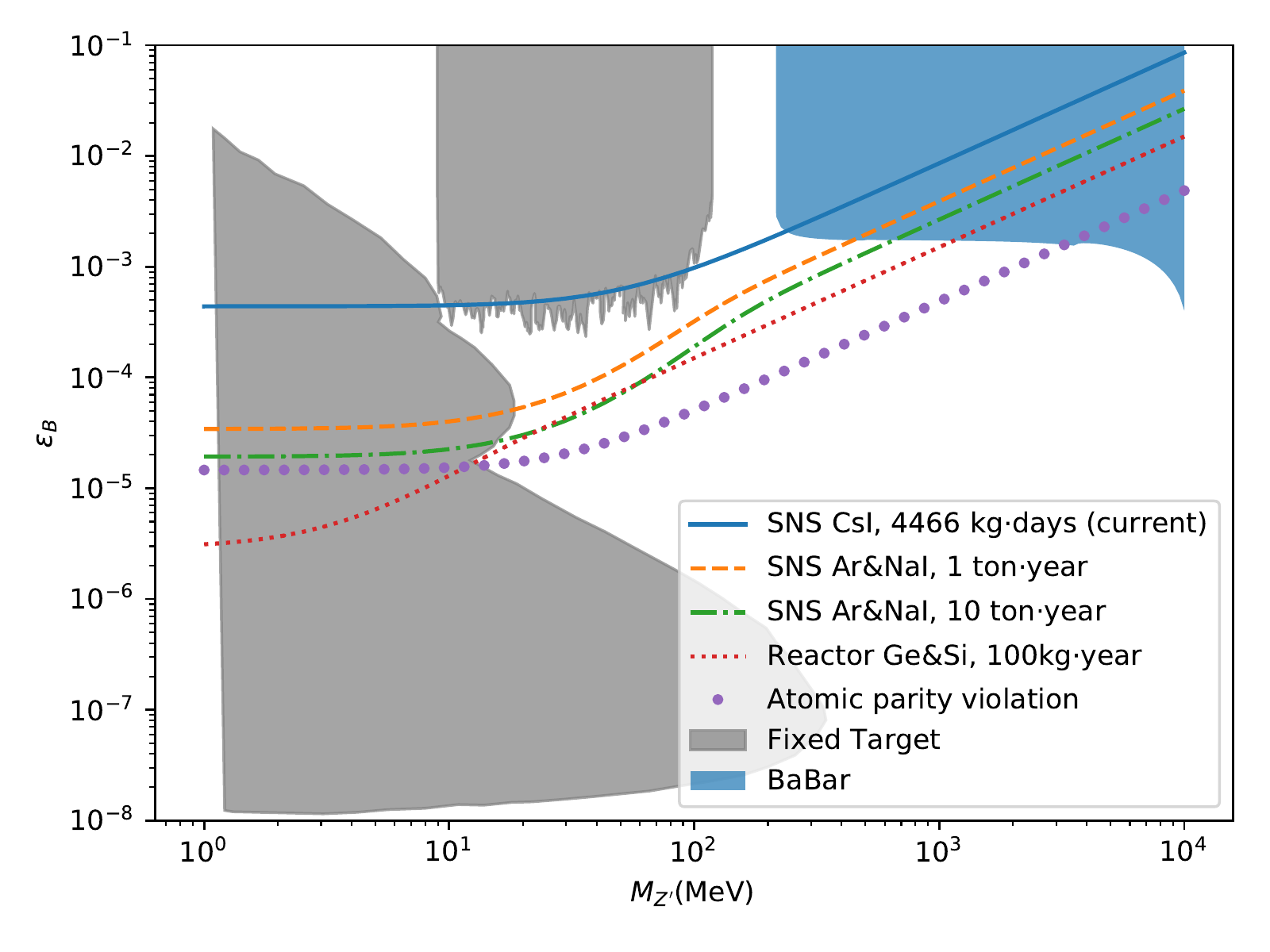}%
\caption{The current and future bounds on the mixing $\epsilon_B$ in the dark hypercharge case are plotted as a function of the $Z'$ mass $M_{Z^\prime}$. The solid blue curve is the current COHERENT limit, the orange dashed and green dot-dashed curves are derived future projections for COHERENT for different luminosities, the red dotted curved is the future projection for a reactor experiment, the purple large-dotted curve is from atomic parity violation, the grey regions are from the NA48/2, E774, E141, and E137 fixed target experiments. The blue shaded region is diallowed by the BaBar results.}
\label{darkb}
\end{figure}

In Fig.~\ref{darkb} we show the COHERENT and reactor reach on $\epsilon_B$ as a function of $M_{Z'}$ using the couplings shown in Eq.(\ref{newcoupling}) and compare it to limits from fixed target, atomic parity violation experiments and the BaBar results. The points where the curves plateau correspond to the energy scale of each experiment above which $M_{Z'}$ dominates over the momentum dependence. In the region allowed by the fixed target experiments, we find that current and projected limits from CE$\nu$NS measurements provide stringent constraints ($10^{-5} < \epsilon < 10^{-2}$) in the mass range $1\,\text{MeV}\, \lesssim m_{Z^\prime} \lesssim 10$ GeV, almost as strong as existing limits from atomic parity violation. The Babar results provide better constraints for $m_{Z^\prime}\sim 10$ GeV. Below about 10 MeV, the future COHERENT constraints are comparable to those from atomic parity violation and reactor experiments are projected to provide stronger limits thanks to the low energies of reactor neutrinos.

\begin{figure}[h]
\includegraphics[scale = .7]{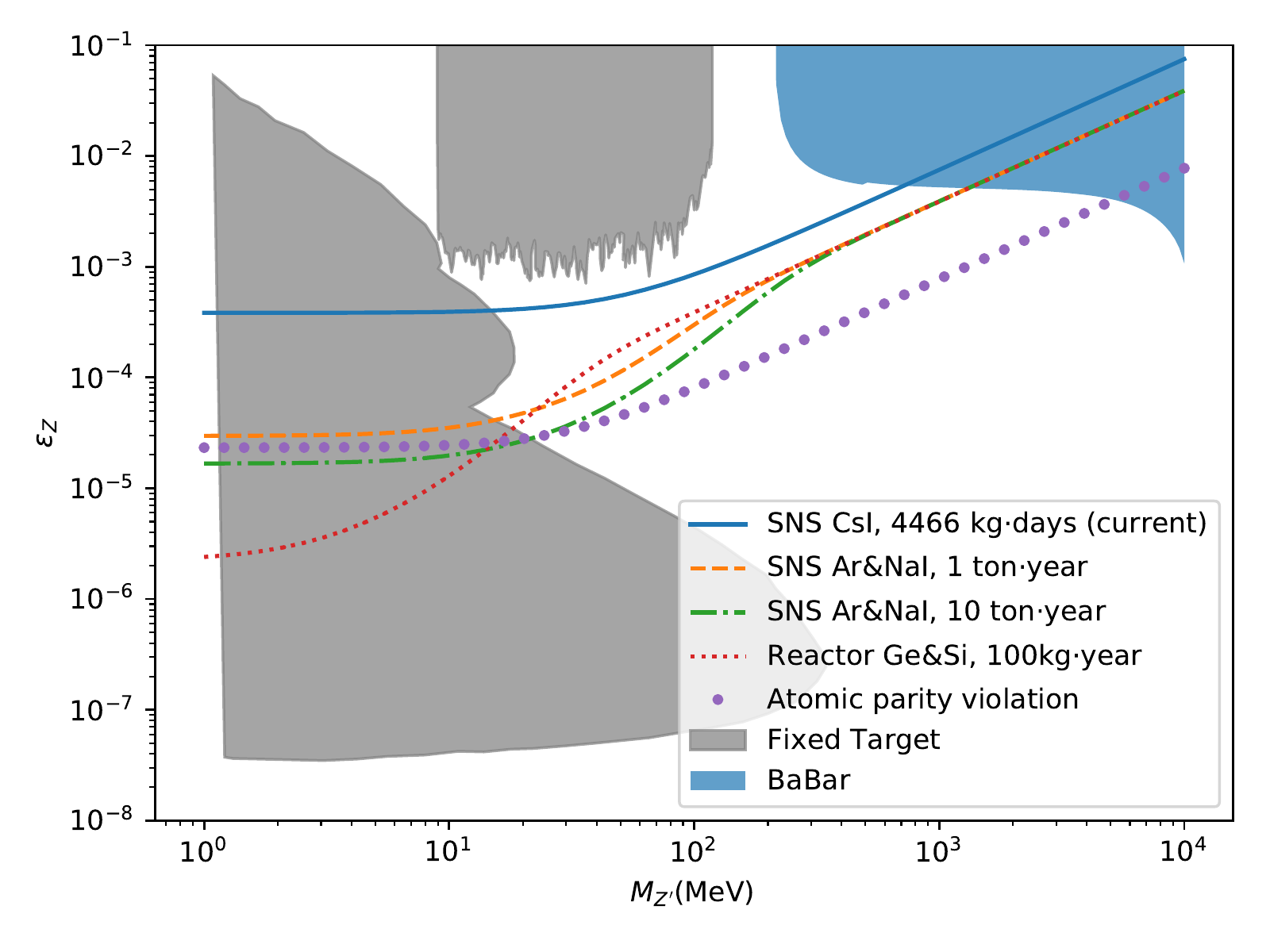}%
\caption{The current and future bounds on the mixing $\epsilon_Z$ in the dark $Z$ case are plotted as a function of the $Z'$ mass $M_{Z^\prime}$. The solid blue curve is the current COHERENT limit, the orange dashed and green dot-dashed curves are derived future projections for COHERENT for different luminosities, the red dotted curved is the future projection for a reactor experiment, the purple large-dotted curve is from atomic parity violation, and the grey regions are from the NA48/2, E774, E141, and E137 fixed target experiments. The blue shaded region is ruled out by the BaBar results.}
\label{darkz}
\end{figure}

In Fig. \ref{darkz}, we show the same constraints applied to $\epsilon_Z$ for the case of a dark $Z$ boson, as given in Eq. (\ref{eq:darkz}). The constraints are similar to the dark hypercharge case with two main differences. First, the window where the \cns constraints start competing with atomic parity violation lies outside the bounds of fixed target experiments. Second, for high values of $m_{Z'}$ the \cns limits become independent of the exposure and detector material. This effect is due to the nature of the $Z'$ coupling as well as the high luminosity compared to the assumed systematic uncertainty. 

In the dark $Z'$ scenario the ratio of BSM to SM couplings to protons and neutrons are identical and equal to $0.27(\epsilon_{Z}/m_{Z^{\prime}})^2$, which limits the distinguishing power of detectors with different proton and neutron content. Coupled with the fact that the number of expected events is sufficiently large for the systematic uncertainty to dominate the statistical uncertainty, this leads to the merger of limits from different detectors and exposures. Note that the same cannot be said about the dark hypercharge scenario. In this case, the relative couplings to protons and neutrons are $-3.3(\epsilon_{Z}/m_{Z^{\prime}})^2$ and $0.06(\epsilon_{Z}/m_{Z^{\prime}})^2$ respectively which significantly enhances the reach when different detectors are combined.

This can also be demonstrated using Eq.~(\ref{chi}). After maximizing the expression with respect to the nuisance parameter $\beta$ and defining $k \equiv \frac{N_{pred}}{N_{bg}+N_{exp}}$, we get:

\begin{align}
\chi^{2} & =\frac{-\left(\sum\frac{N_{pred}N_{exp}}{N_{bg}+N_{exp}}\right)^{2}+\sum\frac{N_{pred}^{2}}{N_{bg}+N_{exp}}\sum\frac{N_{exp}^{2}}{N_{bg}+N_{exp}}+\frac{1}{\sigma^{2}}\sum\frac{\left(N_{pred}-N_{exp}\right)^{2}}{N_{bg}+N_{exp}}}{\frac{1}{\sigma^{2}}+\sum\frac{N_{pred}^{2}}{N_{bg}+N_{exp}}}\\
 & \simeq\sum\frac{N_{exp}^{2}}{N_{bg}+N_{exp}}-\frac{\sum kN_{exp}}{\sum kN_{pred}}\sum kN_{exp}\label{eq:pchisquare}
\end{align}
where in the second line we used the fact that $1/\sigma^2$ is small compared to $N_{exp}$. As mentioned earlier, the $Z'$ coupling relative to the $Z$ coupling is universal in the dark $Z$ scenario which means that $k$ is the same for all detectors and energy bins at the high $M_{Z^{\prime}}$ region. Therefore, Eq. (\ref{eq:pchisquare}) can be solved for $k$ to give 

\begin{equation}
k = \left(\sum\frac{N_{exp}^{2}}{N_{bg}+N_{exp}}-\chi^{2}\right)/\sum N_{exp} \simeq \left(\sum\frac{N_{exp}^{2}}{N_{bg}+N_{exp}}\right)/\sum N_{exp}.
\end{equation} 
In other words, $k$, and hence $\epsilon_Z$ is independent of the exposure. This argument breaks down for the dark hypercharge case due to the detector dependence on $k$.

\begin{figure}[h]
\includegraphics[scale = .7]{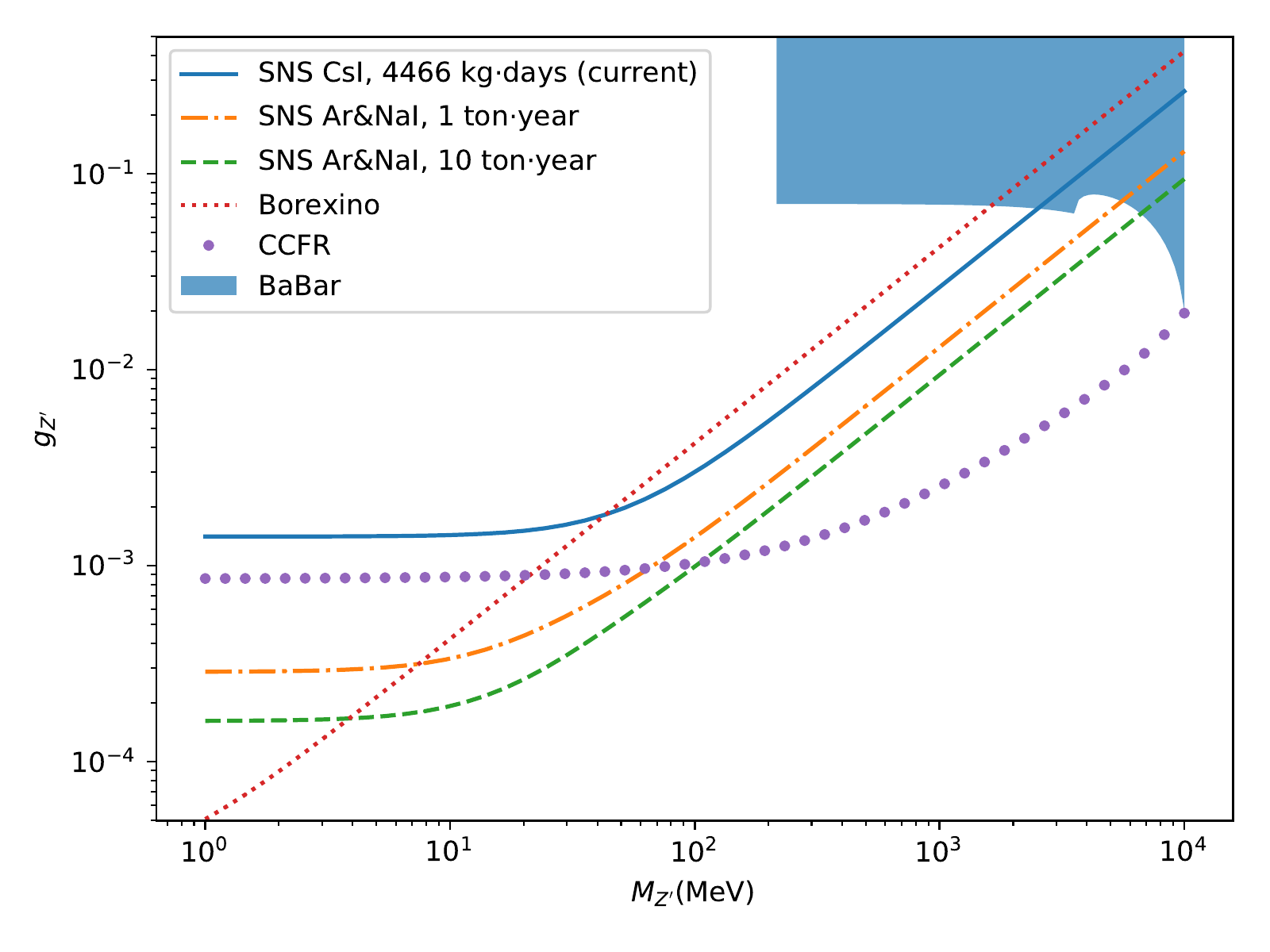}%
\caption{The current and future bounds on the coupling $g_{Z'}$ in the $L_\mu$-$L_\tau$ model  are plotted as a function of the $Z'$ mass $M_{Z^\prime}$. The solid blue curve is the current COHERENT limit, the orange dot-dashed and green dashed are derived future projections for COHERENT for different luminosities, the red dotted curve is from the Borexino measurement of solar neutrinos, and the purple large-dotted curve is from the CCFR measurement of neutrino trident production. The blue shaded region is ruled out by the BaBar results.}\label{lmultau}
\end{figure}

Finally, in Fig.~\ref{lmultau}, we show the limit $g_{Z'}$ as a function of $M_{Z^\prime}$ for the case of $L_\mu$-$L_\tau$ models using current and projected COHERENT results and contrast it against limits from Borexino and CCFR. We find the three experiments to be complementary, and in the mass window $4\,\text{MeV}\, \lesssim m_{Z^\prime} \lesssim 100$ MeV the future COHERENT projections provide the strongest limits ($g_Z'\, \leq \, 1-9 \times 10^{-4}$). Note that the reactor, fixed target, BaBar and atomic parity violation experiments present poor limits in this scenario since they require electron flavor couplings. 

The coherent scattering crossection is dominated by the neutrinos from the $\pi^{\pm}$ decay. The neutrinos can also be produced from the $\pi^0\rightarrow\gamma Z^{\prime}$ decay which subsequently interacts with the target via the SM interaction. The ratio of the  cross-section arising from the second process compared to the first one  can be estimated as $\propto$ ${(1-m^2_{Z^\prime}/m^2_{\pi})^{3/2}}\over{1/(G_F(2 m_N E_r+M^2_{Z^{\prime}}))}$ where $m_N$ is the target mass. The ratio is $\leq 10^{-7}$ for $M_{Z^{\prime}}<0.15$ GeV ($M_{Z^{\prime}}<m_{\pi^0}$) and $E_r<10^{-4}$ GeV (COHERENT experiment). The ratio is similarly suppressed in the case of a bremsstrahlung production of the dark boson. 

\section{Discussion}
\label{sec:discussion}

We have explored the capability of the coherent elastic neutrino-nucleus (CE$\nu$NS) process to explore a light $U(1)$ gauge boson through mixing effects. We looked at three specific scenarios: two scenarios in which the observable couplings to standard model fermions are generated through kinetic mixings associated with the hypercharge boson as well as  mass mixings among $Z$ and $Z^\prime$ arising due to an extended Higgs sector (or via the Stueckelberg mechanism) and one scenario in which the mixing is generated at low energy through standard model fermion loops (the $L_\mu-L_\tau$ scenario). The scenario with a dominant kinetic mixing  is called the dark hypercharge scenario whereas the scenario with a dominant mass mixing is labeled as the dark $Z$ scenario.  We have contrasted the CE$\nu$NS limits for these scenarios with ones from existing experiments.

In the dark hypercharge and dark $Z$ scenarios, we find limits that are complementary to those from fixed target experiments.
 For these scenarios, however, the atomic parity violation experiment provides the best constraint for a large region of parameter space. The Babar results provide better constraints for $m_{Z^\prime}\sim 10$ GeV. In  typical string/M-theory based models, it is argued that $\epsilon\sim 10^{-1}-10^{-3}$~\cite{Acharya:2016fge} where hypercharge mixing is considered. Based on Fig. 2, we can see that most of the parameter space for these models is ruled out. However, it is also argued in the context of LVS (Large Volume Scenario) that $\epsilon\sim 10^{-6}-10^{-8}$, which survives the experimental constraint. In such models, the dark matter candidate  in the visible sector decays into hidden sector particles via 2 and 3 body decay modes and the lifetime  depends on $\epsilon$. The restriction on $\epsilon$ from our analysis  constrains the lifetime of the dark matter particle into 3-body and 2-body decay modes  to be $10^{-10}-10^{-13}$ sec and $10^{-18}-10^{-21}$ sec, respectively, where these ranges correspond to a $Z^{\prime}$ mass spanning $10^{-3}-10$ GeV.

In the $L_\mu-L_\tau$ scenario, the COHERENT sensitivity is projected to dominate over those from Borexino and CCFR in the mass range $4\,\text{MeV}\, \lesssim m_{Z^\prime} \lesssim 100$ MeV. Note that in this case the limits reactors provide are very weak due to the lack of direct coupling to electron flavor neutrinos. This observation emphasizes the value of experiments such as COHERENT in probing new physics despite the lower degree of coherency in the neutrino source. 

The \cns experimental program, including the COHERENT experimental effort, is very new  in its development, and there is a lot of room for improvement in detector technology and control of systematics. In addition to the detector technologies that we have discussed, new low threshold Ge detectors will soon be deployed for \cns detection at COHERENT \cite{collar}. The same can be said regarding reactor experiments of which we only studied the case of a low yield (1 MW) reactor at a relatively short distance (1 m) from the detector. A high yield (1 GW) reactor at a 10 m distance would provide a less stable neutrino source but a 10 times larger rate. The models we have studied demonstrate the need to utilize such a wide variety of neutrino sources and detection techniques.

\section{Acknowledgments}

BD and  LES acknowledge support from DOE Grant de-sc0010813. GLK acknowledges support from
grant DE-SC0007859. SL acknowledges support from NSF grant PHY-1522717.
JBD would like to thank the Mitchell Institute for Fundamental Physics and Astronomy at 
Texas A\&M University for their generous hospitality, and MA would like to thank the 
Mitchell Institute for support. BD, GLK and LES would like to thank M. Perry, C. Pope and A. Zytkow for organizing the Cambridge -Mitchell Workshop where this work was initiated. BD  thanks M. Perry for discussions.

\bibliography{epsilon}

\end{document}